\documentclass[conference,a4paper]{IEEEtran}
\usepackage{algorithm}
\usepackage{algpseudocode}
\usepackage{amsfonts}
\usepackage{bm}
\usepackage{amsmath}
\usepackage{cite}
\usepackage{stfloats}
\usepackage{amssymb}
\usepackage{enumerate}
\usepackage{color,xcolor}
\usepackage{graphicx}
\usepackage{multirow,tabularx}
\usepackage{subfigure}
\usepackage{hyperref}
\usepackage[T1]{fontenc}
\usepackage{flushend}
\usepackage{mathrsfs}
\usepackage{float}

\makeatletter

\newcommand{\Rmnum}[1]{\expandafter\@slowromancap\romannumeral #1@}
\makeatother

\ifCLASSINFOpdf

\else

\fi

\begin{document}
\title{6G Downlink Transmission via Rate Splitting Space Division Multiple Access Based on Grouped Code Index Modulation}
\author{\IEEEauthorblockN{Wenchao Zhai$^{1}$, Yishan Wu$^{1}$, Jun Zhao$^2$, \emph{Member, IEEE}, Huimei Han$^3$}

\vspace{1.5mm}
\\
\fontsize{10}{10}\selectfont\itshape
$^1$Key Laboratory of Electromagnetic Wave and Metrology of Zhejiang Province, \\College of Information Engineering, China Jiliang University, Hangzhou, China\\
$^2$School of Computer Science and Engineering, Nanyang Technological University, Singapore\\
$^3$ College of Information Engineering, Zhejiang University of Technology,
Hangzhou, China\vspace{1.5mm}\\

\fontsize{9}{9}
$^1$\{zhaiwenchao@cjlu.edu.cn,\,1249130501@qq.com\}, $^2$junzhao@ntu.edu.sg, $^3$hmhan1215@zjut.edu.cn\vspace{-2mm}
}


%

\maketitle
\begin{abstract}
A novel rate splitting space division multiple access (SDMA) scheme based on grouped code index modulation (GrCIM) is proposed for the sixth generation (6G) downlink transmission. The proposed RSMA-GrCIM scheme transmits information to multiple user equipments (UEs) through the space division multiple access (SDMA) technique, and exploits code index modulation for rate splitting. Since the CIM scheme conveys information bits via the index of the selected Walsh code and binary phase shift keying (BPSK) signal, our RSMA scheme transmits the private messages of each user through the indices, and the common messages via the BPSK signal. Moreover, the Walsh code set is grouped into several orthogonal subsets to eliminate the interference from other users. A maximum likelihood (ML) detector is used to recovery the source bits, and a mathematical analysis is provided for the upper bound bit error ratio (BER) of each user. Comparisons are also made between our proposed scheme and the traditional SDMA scheme in spectrum utilization, number of available UEs, etc. Numerical results are given to verify the effectiveness of the proposed SDMA-GrCIM scheme.
\end{abstract}
\vbox{}
\IEEEpeerreviewmaketitle
\section{Introduction}
\IEEEPARstart{W}{ith} the commercialization of the fifth generation (5G) network in 2020~\cite{5G2020Yang,5G2020Storck}, massive connectivity is required due to the huge numbers of user equipment (UE) in the network. It is predicted that the global UEs will grow to 13.1 billion by 2023~\cite{5G2020Li}, and the ``ubiquitous connectivity'' is expected after ten years in the sixth generation (6G) network~\cite{6G2020Liu,6G2020Han,6G2020Zhao,6G2021Han}. The explosive growth of numbers of UEs makes the sustainable development of 6G network a challenging task, and space division multiplexing access (SDMA) can be used to satisfy the requirement of ``ubiquitous connectivity''~\cite{SDMA2020Zhang,SDMA2020Vidal,SDMA2020Chen}. In SDMA, multiple UEs are accessed to the network through a linear precoding operation, and the signals from other UEs are treated as interference. However, SDMA is more suitable for unicast service transmission. To apply SDMA technology in broadcast or multicast services, a concept of rate splitting (RS) is introduced, where the transmitted messages of each UE are linearly precoded and split into common and private messages~\cite{RSMA2020Clerckx,RSMA2020Yin,RSMA2020Naser}.

However, in classical RS multiple accessing (RSMA), multiple common and private messages are transmitted by the base station (BS) after superposition~\cite{5G2020Li,RSMA2020Jaafar,RSMA2020Dizdar}. At the receiver, successive interference cancellation is exploited to decode the common messages and private messages subsequently. Therefore, high interference introduced from the private messages exists when decoding the common messages. To guarantee the signal to interference and noise ratio (SINR) at the receiver, various resource allocation methods are investigated. However, to the best of our knowledge, all these methods aim at the alleviation instead of elimination of the interferences, whether the inter-user interference or the interference between common and private messages of a UE.

We will propose a novel RSMA scheme based on grouped code index modulation (GrCIM) scheme to realize the accessibility of multiple UEs and the rate splitting. IM technology conveys information bits through two parts, one part for the selected resource entities, such as time slot, subcarriers, antennas, etc., and the other part for classical amplitude phase modulation (APM)~\cite{IM2019Mao}. Since the information transmission for resource entity selections doesn't occupy any power and spectrum resources, IM technology has both high power and high spectrum efficiencies, which is attractive for future 6G networks. Kaddoum \emph{et al.}~\cite{CIM2015Kaddoum} first propose the concept of CIM, which transmits the index bits information via the selection of the spreading code from a given code set, and the binary phase shift keying (BPSK) signals, seperatly. Generally, the Walsh codes are exploited for spreading spectrum~\cite{CIM2015Kaddoum1}. In our proposed RSMA scheme, we group the Walsh code set into several orthogonal subsets, and each subset for a specific UE. Therefore, inter-user interference can be eliminated due to the orthogonality between different Walsh codes. On the other hand, by applying the common and private messages of a UE to the two parts of a CIM signal, i.e., index selection information and BPSK signals, interference between common and private messages can also be avoided.

The remainder of this paper is organized as follows. System model and the principle of the RSMA-GrCIM scheme are described in detail in Section II. Section III explains the maximum likelihood (ML) algorithm in detail. The performance analysis is presented in Section IV, including the upper bound bit error ratio (BER) of the RSMA-GrCIM scheme, and the spectrum utilization, etc. Numerical results are presented in Section V to verity our analysis, and Section VI concludes this paper.

\emph{Notation:} We explain the notations used throughout our manuscript as follows. Bold letters of lower and upper case denote vector and matrix, respectively. $\lfloor \cdot \rfloor$ denotes the largest integer that is no greater than the argument. $\mathbb P[\cdot]$ represents the probability of an event. The subscripts $(\cdot)^T$, $(\cdot)^*$, and $(\cdot)^H$ represent the operation of transpose, conjugate, and Hermitian transpose, respectively. $||\cdot||$ means the Frobenius norm of a matrix or vector. $\mathbb E(\cdot)$ is the expectation operation. $\textbf{I}_N$ represents an $N$-by-$N$ identity matrix, and $(\cdot) \in \mathbb C^{m\times n}$ means the argument is an $m\times n$ matrix. $\Re(\cdot)$ and $\Im(\cdot)$ are the real and imaginary part of the argument.

\section{System Model}
Consider a downlink transmission SDMA scheme where $N_u$ single antenna UEs communicate with a BS with $N_T$ transmit antennas. A Hadamard code matrix with the size of $L_c \times L_c$ and the element of $\pm 1$ is used for GrCIM. Therefore, the code set has $L_c$ spreading codes with each code of length $L_c$. To guarantee the orthogonality between the UEs, the Walsh code set is grouped into $N_u$ subsets. To obtain the target of rate splitting, the source information bits transmitted to the $k$th UE are divided into two parts: index selections of the Walsh codes for private messages, and the BPSK signals for common messages.


We denote $s_{k,l_I}$ and $s_{k,l_Q}$ ($l_I,l_Q\in\mathcal L=\left\{0,1\right\}$) as the BPSK signals of in-phase and quadrature component, respectively. Therefore, the modulation alphabet is $M=2$. The values of $s_{k,l_I}$ and $s_{k,l_Q}$ are taken from the constellation set $\mathcal{M}=\left\{1/\sqrt{2}, -1/\sqrt{2}\right\}$ to ensure that the transmitted common and private messages have unit power. $\mathcal{S}_k$ ($k=1,2,...,N_u$) represents the Walsh code set for the $k$th UE, with $\mathcal N=\left\{0,1,...,N_c-1\right\}$ denotes the Walsh code index set.  $\textbf{c}^{H}_{k,n_I}\in \mathbb C^{1\times Lc}$ and $\textbf{c}^{H}_{k,n_Q}\in \mathbb C^{1\times Lc}$ ($n_I,n_Q\in\mathcal N$) represent the selected Walsh codes used for spectrum spreading of in-phase and quadrature signals, respectively.

Suppose there are $2m$ bits conveyed for each UE's private messages, the number of Walsh codes candidate is $N_c=2^m$. The data rates in terms of bits per channel use (bpcu) for the UE and the BS are expressed as
\begin{equation}\label{RUE}
R_\text{UE} = 2m+2 \text{(bpcu)} ,
\end{equation}
and
\begin{equation}\label{RBS}
2mN_u+2 \le R_\text{BS} \le N_u(2m+2) \text{(bpcu)} ,
\end{equation}
respectively. And the minimum/maximum transmission rate is achieved in the case of broadcast/unicast scenario.

The transmitted signal of the BS is given by
\begin{equation}\label{send}
\textbf{x} = \sum_{k=0}^{N_u}{\textbf{x}_k},
\end{equation}
with $\textbf{x}_k\in \mathbb C^{N_T\times 1}$ (k=1,...,$N_u$) being the signal transmitted to the $k$th UE
\begin{equation}\label{sendUE}
\textbf{x}_k=\textbf{w}_k(\textbf{c}^{H}_{k,n_I}s_{k,l_I}+j\textbf{c}^{H}_{k,n_Q}s_{k,l_Q}),
\end{equation}
where $\textbf{w}_k \in \mathbb C^{N_T\times 1}$ is the $k$th UE's precoding vector, and $j=\sqrt{-1}$ is the imaginary unit.

After passing through a Rayleigh fading channel, each UE can receive the transmitted signal. Suppose perfect synchronization is conducted, the expression of the received signal for the $k$th UE is
\begin{equation}\label{receive}
\begin{array}{lcl}
\textbf{y}_k^H&=&\textbf{h}_k^H\textbf{x} \\
&=&\textbf{h}_k^H\textbf{x}_k+\sum_{i=1,i\ne k}^{N_u}\textbf{h}_k\textbf{x}_i+\textbf{n}_k^H,
\end{array}
\end{equation}
where $\textbf{y}_k^H\in \mathbb C^{1\times L_c}$ is the $k$th UE's received CIM signal, $\textbf{h}_k^H\in\mathbb C^{1\times N_T}$ is the Rayleigh fading channel vector between the BS and the $k$th UE whose entries are independent and identically distributed (iid) complex Gaussian random variables (RVs) with variance $\sigma^2_k$,  and $\textbf{n}_k^H\in \mathbb C^{1\times L_c}$ is the corresponding noise vector whose elements are iid complex Gaussian RVs of variance $N_0$.

In our scheme, we use zero forcing (ZF) method to design the precoding vector $\textbf{w}_k$ (k=1,...,$N_u$), which eliminates the inter-antenna interference~\cite{RSM2011Yang}. Therefore, the precoding vector $\textbf{w}_k$ can be expressed as
\begin{equation}\label{precoding}
\textbf{w}_k=\dfrac{\beta_k \textbf{h}_k}{||\textbf{h}_k||^2},
\end{equation}
where $\beta_k$ is used for power allocation. Suppose the power of each UE's transmitted signal is denoted as $P_k$. Then we have
\begin{equation}\label{powerUE}
P_k=\dfrac{\beta^2_k}{||\textbf{h}_k||^2}.
\end{equation}

To guarantee that the transmitted signals have unit power, the power of all UEs is constrained by
\begin{equation}\label{power}
\sum_{k=0}^{N_u} P_k = 1.
\end{equation}

On the other hand, to ensure fairness, the UEs with large channel fading attenuation should be allocated to more power resource. Therefore, for the $k$th UE, the allocated power should be inversely proportional to the variance of its channel fading vector, i.e.,
\begin{equation}\label{power-variance}
P_k \propto \dfrac{1}{\sigma^2_k}.
\end{equation}

From equations. (\ref{powerUE})-(\ref{power-variance}), we obtain
\begin{subequations}
\begin{align}
P_k=\dfrac{P}{\sigma^2_k},\label{powerUEk} \\
\beta_k=\sqrt{\dfrac{P||\textbf{h}_k||^2}{\sigma^2_k}}, \label{betaUEk}
\end{align}
\end{subequations}
where
\begin{equation}\label{power-coeff}
P=\dfrac{1}{\sum_{i=1}^{N_u}{1/\sigma^2_i}}
\end{equation}
is the normalized coefficient. In the next section, we will demonstrate the fairness of the power allocation method.

Substituting Eqs. (\ref{send}) and (\ref{precoding}) into (\ref{receive}), we get the expression of the $k$th UE's received signal:
\begin{equation}\label{receiveUEk}
\textbf{y}_k =\beta_k(\textbf{c}^{H}_{k,n_I}s_{k,l_I}+j\textbf{c}^{H}_{k,n_Q}s_{k,l_Q})+\sum_{i=1,i\ne k}^{N_u}\textbf{h}_k\textbf{x}_i+\textbf{n}_k^H.
\end{equation}

\section{ML Detector}
Eq. (\ref{receiveUEk}) gives the received signal expression of the $k$th UE. The second term in the equation is the interference from the other UEs. However, as the Walsh codes are selected from different code sets, the inter-user interference can be eliminated by exploiting the orthogonality between different Walsh codes, i.e.,
\begin{equation}\label{walsh-cor}
\textbf{c}_i^H\textbf{c}_k=\left\{\begin{array}{rr}L_c & i=k \\ 0 & i \ne k \end{array} \right. .
\end{equation}

Therefore, for the $k$th UE, the received signal can be sent into multiple correlators, and the correlated signals of the in-phase part are expressed as
\begin{equation}\label{cor}
\renewcommand\arraystretch{2}
\begin{array}{lcl}
r_{k,\tilde n_I}&=&\Re[{\textbf{y}_k}]\textbf{c}_{k,\widetilde n_I} \\
&=&\beta_k[\textbf{c}^{H}_{k,n_I}\textbf{c}_{k,\widetilde n_I}s_{k,l_I}]+ n_{k,\tilde n_I},
\end{array}
\end{equation}
where $\textbf{c}_{k,\widetilde n_I}$, ($\widetilde n_I\in \mathcal N$) $\in \mathcal{S}_k$ is the possible transmitted Walsh code of in-phase component; and $n_{k,\tilde n_I}$ is the noise term after passing through the correlator:
\begin{equation}\label{noise-corI}
n_{k,\tilde n_I}=\Re[\textbf{n}_k^H]\textbf{c}_{k,\widetilde n_I}.
\end{equation}

Similarly, the quadrature signals after passing through the correlators are written as
\begin{equation}\label{cor}
\renewcommand\arraystretch{2}
\begin{array}{lcl}
r_{k,\tilde n_Q}&=&\Im[{\textbf{y}_k}]\textbf{c}_{k,\widetilde n_Q} \\
&=&\beta_k[\textbf{c}^{H}_{k,n_Q}\textbf{c}_{k,\widetilde n_Q}s_{k,l_Q}]+ n_{k,\tilde n_I},
\end{array}
\end{equation}
with $\textbf{c}_{k,\widetilde n_Q}$, ($\widetilde n_Q\in \mathcal N$) $\in \mathcal{S}_k$ and
\begin{equation}\label{noise-corQ}
n_{k,\tilde n_Q}=\Im[\textbf{n}_k^H]\textbf{c}_{k,\widetilde n_Q}.
\end{equation}

Because the BPSK signals of in-phase and quadrature component are independent, we can decouple the real and imaginary part and process them separately with the same operation. For convenience, only the in-phase part is utilized for analysis, and the same result can be obtained for the quadrature part as well.

Substituting Eq. (\ref{walsh-cor}) into (\ref{cor}), we obtain
\begin{equation}\label{receive-walsh}
r_{k,\tilde n_I}=\left\{\begin{array}{ll} \beta_kL_cs_{k,l_I}+n_{k,\tilde n_I}& \tilde n_I =n_I\\ n_{k,\tilde n_I}& \tilde n_I \ne n_I \end{array} \right.
\end{equation}

\begin{figure*}[ht]
  \centering
  \includegraphics[scale=0.7]{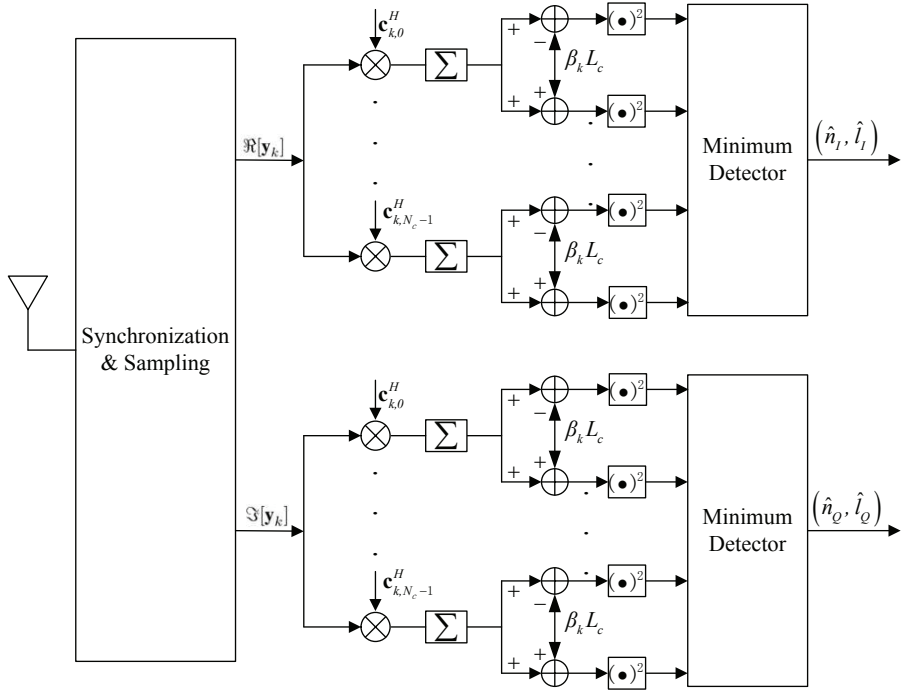}\\
  \caption{Structure of each UE's ML detector.}\label{ML}
\end{figure*}

From Eq. (\ref{receive-walsh}), we get the ML algorithm:
\begin{subequations}\label{search}
\begin{align}
[\hat{n}_I, \hat{l}_I]=\mathop {\arg \min }\limits_{\tilde n_{\tilde n_I}\in\mathcal N, \tilde l_I\in\mathcal L}(r_{k,\tilde n_I}-\beta_kL_cs_{k,\tilde l_I})^2, \label{ML-I} \\
[\hat{n}_Q, \hat{l}_Q]=\mathop {\arg \min }\limits_{\tilde n_{\tilde n_Q}\in\mathcal N, \tilde l_Q\in\mathcal L}(r_{k,\tilde n_Q}-\beta_kL_cs_{k,\tilde l_Q})^2. \label{ML-Q}
\end{align}
\end{subequations}
Thereafter, we can achieve the ML detector for each UE according to Eq. (\ref{receive-walsh}), and the corresponding structure is depicted in Fig.~\ref{ML}.

\section{Performance Analysis}
In this section, we will give performance analysis of our proposed rate-splitting SDMA-GrCIM scheme, including the upper bound BER and the comparisons between our scheme and the traditional SDMA scheme in spectrum utilization, maximum number of available UEs, etc.
\subsection{Upper Bound of BER Performance}
To achieve the BER performance of our proposed SDMA-GrCIM scheme, the pair-wise error probability (PEP) is used for theoretical analysis. Suppose $(n_I,l_I)$ is transmitted as the private and common messages, an error occurs when $(\tilde n_I,\tilde l_I)\ne(n_I,l_I)$. We denote this event as $(n_I,l_I)\rightarrow(\tilde n_I,\tilde l_I)$. Therefore, the PEP can be calculated as
\begin{equation}\label{PEP}
\renewcommand\arraystretch{2}
\begin{array}{l}
\mathbb P[(n_I,l_I)\rightarrow (\tilde n_I,\tilde l_I)]=\\
{\kern 15pt} \mathbb P[(r_{k,l_I}-\beta_kL_cs_{k,l_I})^2 \ge (r_{k,\tilde n_I}-\beta_kL_cs_{k,\tilde l_I})^2].\\
\end{array}
\end{equation}

It is obvious that
\begin{equation}\label{nkI}
(r_{k,l_I}-\beta_kL_cs_{k,l_I})^2=(n_{k,n_I})^2,
\end{equation}
and substituting Eq. (\ref{receive-walsh}) into (\ref{PEP}), we get
\begin{equation}\label{PEP1}
\renewcommand\arraystretch{2}
\begin{array}{l}
(r_{k,\tilde n_I}-\beta_kL_cs_{k,\tilde l_I})^2=\\
{\kern 10pt}\left\{ \begin{array}{ll} (n_{k,\tilde n_I}-\beta_kL_cs_{k,\tilde l_I})^2 &n_I\ne \tilde n_I \\ (\beta_kL_cs_{k,\tilde l_I}-\beta_kL_cs_{k,l_I}+n_{k,\tilde l_I})^2 & n_I=\tilde n_I,l_I\ne\tilde l_I\end{array} \right.
\end{array}
\end{equation}

After arranging Eqs. (\ref{nkI}) and (\ref{PEP1}), we can obtain
\begin{equation}\label{PEP2}
\renewcommand\arraystretch{2}
\begin{array}{ll}
\mathbb P[(n_I,l_I)\rightarrow (\tilde n_I,\tilde l_I)] = \\
\left\{\begin{array}{ll} \mathbb P[X_1 \ge \beta_k^2L_c^2s_{k,\tilde l_I}^2] & n_I\ne\tilde n_I, \\ \mathbb P[X_2\ge\beta_k^2L_c^2(s_{k,\tilde l_I}-s_{k,l_I})^2] & n_I=\tilde n_I,l_I\ne\tilde l_I, \end{array} \right.
\end{array}
\end{equation}
where $X_1=2\beta_kL_cs_{k,\tilde l_I}n_{k,\tilde n_I}+\Delta n$ and $X_2=2\beta_kL_c(s_{k,\tilde l_I}-s_{k,l_I})n_{k,\tilde n_I}+\Delta n$ are RVs and $\Delta n=(n_{k,n_I})^2-(n_{k,\tilde n_I})^2$.

To achieve the closed-form expression of PEP, it is required to get the probability distribution of RVs $X_1$ and $X_2$. The exact derivation of the probability distribution is involved because of the square term of Gaussian RV in $\Delta_n$. However, as the transmitted CIM signals for each UE is of unit power, the term $\Delta_n$ can be omitted in high signal to noise ratio (SNR). Therefore, $X_1$ and $X_2$ are approximated as mean zero Gaussian RVs in high SNR region. From Eq. (\ref{noise-corI}) we can obviously see that the variance of $n_{k,\tilde n_I}$ is $\sigma^2_{k,I}=L_cN_0/4$, and the variances of $X_1$ and $X_2$ are
\begin{subequations}\label{var-X}
\begin{align}
\sigma^2_{X_1}=\beta_k^2L_c^3N_0s_{k,\tilde l_I}^2 \label{var-X1},\\
\sigma^2_{X_2}=\beta_k^2L_c^3N_0(s_{k,\tilde l_I}-s_{k,l_I})^2 \label{var-X2}.
\end{align}
\end{subequations}
Therefore, $X_1$ and $X_2$ are Gaussian distributed RVs with respective variances $\sigma^2_{X_1}$ and $\sigma^2_{X_2}$.

Substituting Eq. (\ref{var-X}) into (\ref{PEP2}), the PEP can be calculated under the condition of $\beta_k$:
\begin{equation}\label{PEPI}
\begin{array}{l}
\mathbb P[(n_I,l_I)\rightarrow (\tilde n_I,\tilde l_I)|\beta_k]\approx\\
{\kern 40pt}\left\{\begin{array}{ll}Q\left(\sqrt{\left(\dfrac{\beta_k^2L_cE_c}{2N_0}\right)}\right) & n_I\ne\tilde n_I, \\ Q\left(\sqrt{\left(\dfrac{2\beta_k^2L_cE_c}{N_0}\right)}\right) & n_I=\tilde n_I,l_I\ne\tilde l_I, \end{array} \right.
\end{array}
\end{equation}
where the relations $s_{k,\tilde l_I}^2=1/2$, $(s_{k,\tilde l_I}-s_{k,l_I})^2=2$ and $N_0=1/(E_c/N_0)$ are used when calculating the final results of PEP, with $E_c$ being the energy of the spreading chip. The $Q()$ function is expressed as
\begin{equation}\label{Qfun}
Q(x)=\dfrac{1}{\sqrt{2\pi}}\int_x^{+\infty}e^{-\frac{t^2}{2}}dt.
\end{equation}


The same result can be obtained for the quadrature part as well. Since the PEPs for in-phase and quadrature components have the same expression, the upper bound BER for the whole complex signals is the same with that for the in-phase/quadrature signals.

Since the power factor $\beta_k$ is dependent to the channel fading vector as illustrated in Eq. (\ref{betaUEk}), the average BER upper bound of SDMA-GrCIM for the $k$th UE under the channel fading condition can be expressed as
\begin{equation}\label{Pb}
\begin{array}{l}
P_{b,k}|\textbf{h}^H_k \le\dfrac{1}{MN_c}\sum\limits_{\scriptstyle{n_I} \in\mathcal N\hfill\atop
{\scriptstyle{{\tilde n}_I} \in \mathcal N\hfill\atop
\scriptstyle{n_I} \ne {{\tilde n}_I}\hfill}}\sum\limits_{\scriptstyle{l_I} \in \mathcal L\hfill\atop
\scriptstyle{{\tilde l}_I} \in \mathcal L\hfill}\dfrac{N_b(n_I\rightarrow\tilde n_I)}{m}Q\left(\sqrt{\dfrac{\beta_k^2L_cE_c}{2N_0}}\right)\\
{\kern 30pt}+\dfrac{1}{MN_c}\sum\limits_{\scriptstyle{n_I} \in\mathcal N\hfill}\sum\limits_{\scriptstyle{l_I} \in \mathcal L\hfill\atop
{\scriptstyle{{\tilde l}_I} \in \mathcal L\hfill\atop\scriptstyle{l_I}\ne{\tilde l_I}\hfill}}{N_b(l_I\rightarrow\tilde l_I)}Q\left(\sqrt{\dfrac{2\beta_k^2L_cE_c}{N_0}}\right),
\end{array}
\end{equation}
where $N_b(n_I\rightarrow\tilde n_I)$ and $N_b(l_I\rightarrow\tilde l_I)$ are the number of error bits when $n_I\ne\tilde n_I$ and $l_I\ne\tilde l_I$. Therefore, it follows that $N_b(l_I\rightarrow\tilde l_I)=1$ for BPSK signals. Eq. (\ref{Pb}) can be further simplified as
\begin{equation}\label{Pb1}
\begin{array}{ll}
P_{b,k}|\textbf{h}^H_k &\le \dfrac{MN_c}{2}Q\left(\sqrt{\dfrac{\beta_k^2L_cE_c}{2N_0}}\right)+Q\left(\sqrt{\dfrac{2\beta_k^2L_cE_c}{N_0}}\right)\\
&=N_cQ\left(\sqrt{\dfrac{\beta_k^2L_cE_c}{2N_0}}\right)+Q\left(\sqrt{\dfrac{2\beta_k^2L_cE_c}{N_0}}\right).
\end{array}
\end{equation}
Thus, the upper bound BER is represented as
\begin{equation}\label{Pb2}
P_{b,k}\le N_c\mathbb E\left\{Q\left(\sqrt{\dfrac{\beta_k^2L_cE_c}{2N_0}}\right)\right\}+\mathbb E\left\{Q\left(\sqrt{\dfrac{2\beta_k^2L_cE_c}{N_0}}\right)\right\}.
\end{equation}
The closed-form upper bound BER can be achieved using the conclusion in \cite{Digital2001Proakis}:
\begin{equation}\label{Pb-final}
P_{b,K}=N_c\Omega_1+\Omega_2,
\end{equation}
with
\begin{equation}\label{omega}
\Omega_i=\left(\dfrac{1-\mu_i}{2}\right)^{N_T}\sum\limits_{n=0}^{N_T-1}\left(\begin{array}{c}N_T-1+n\\ n \end{array} \right)\left(\dfrac{1+\mu_i}{2}\right)^{n},
\end{equation}
and
\begin{equation}\label{mu}
\mu_i=\sqrt{\dfrac{\gamma_i}{1+\gamma_i}}, i=1,2,
\end{equation}

\begin{equation}\label{gamma}
\gamma_1=\dfrac{\gamma_2}{4}, \gamma_2=\dfrac{PL_cE_c}{N_0}.
\end{equation}

From eqs. (\ref{Pb-final})-(\ref{gamma}), it is known that the $k$th UE's upper bound BER has no relation with its channel fading variance $\sigma_k^2$, which verifies the fairness of the power allocation method.

\subsection{Comparisons Between SDMA-GrCIM and Traditional SDMA}

\subsubsection{Spectrum Utilization}For our proposed SDMA-GrCIM scheme, the transmission rate for each UE must be the same and is calculated as $R_\text{UE}=(2m+2)$bpcu, with 2m bits for private messages and 2 bits for common messages. And in the perspective of the BS, the transmission rate varies from $(2mN_u+2)$bpcu to $N_u(2m+2)$bpcu according to different transmission scenarios as illustrated in Eq. (\ref{RBS}). However, due to the spreading spectrum in our scheme, the expressions for the respective spectrum utilization for the UE and BS are
\begin{equation}
\Lambda_\text{UE}^\text{GrCIM}=(2m+2)/L_c  (\text{bps/Hz}),
\end{equation}
and
\begin{equation}
\Lambda_\text{BS}^\text{GrCIM}= R_\text{BS}/L_c (\text{bps/Hz}).
\end{equation}

For the traditional SDMA scheme, the transmission rate for each UE can be different. Suppose only the APM signals are used for the transmission of private and common messages, and the corresponding bits numbers are $m_{k,p}$ and $m_{k,c}$, the expressions of the spectrum utilization for the UE and BS are
\begin{equation}
\Lambda_\text{UE}^\text{SDMA}=m_{k,p}+m_{k,c} (\text{bps/Hz}),
\end{equation}
and
\begin{equation}
\Lambda_\text{BS}^\text{SDMA}=\sum\limits_{k=1}{N_u}(m_{k,p}+m_{k,c})(\text{bps/Hz}).
\end{equation}

In general, the spectrum utilization of our proposed SDMA-GrCIM scheme is much lower than its classical counterpart due to the spreading spectrum.

\subsubsection{Received SINR}\label{received-SINR}
The energy between the transmitted CIM signals $E_c$ and the source information bits $E_b$ is $E_c=E_b\Lambda_\text{BS}^\text{GrCIM}$. There exists no inter-user interference due to the orthogonality between different Walsh codes, and the SINR is proportional to $L_cE_c$ as illustrated in Eq. (\ref{Pb2}). Since $L_cE_c=R_\text{BS}E_b$ and $R_\text{BS}$ is approximately proportional to $R_\text{BS}$, the received SINR for a specific UE increases with the number of available Walsh code in the code subset. Larger $N_c$ requires more information bits to choose the index of CIM, and thus leads to higher transmission rate of the BS.

For the traditional SDMA, both the common message $x_{k,c}$ and private message $x_{k,p}$ are precoded to be transmitted to the UE~\cite{5G2020Li,SDMA2020Zhang,RSMA2020Yin}:
\begin{equation}
\textbf{x}^\text{SDMA}=\sum\limits_{k=1}^{N_u}\textbf{w}_{k,c}^\text{SDMA}x_{k,c}+\textbf{w}_{k,p}^\text{SDMA}x_{k,p},
\end{equation}
and the inter-user interference can hardly be eliminated completely, and the SINR will degrade with the number of UE increases.

\subsubsection{Maximum Number of UEs}
For the sake of orthogonality between different UEs, the total number of Walsh codes should not be greater than $L_c$. Therefore, $N_uN_c \le L_c$ is required. The maximum number of UEs is $N_{u,max}^\text{GrCIM}=L_c$. In such a case, no private messages are conveyed and CIM changes to code division multiple access (CDMA).

For the classical SDMA, the maximum number of beams for $N_T$ transmit antennas is $N_T$. Therefore, at most $N_T$ UEs can be accessed to the BS simultaneously, i.e., $N_{u,max}^\text{SDMA}=N_T$.

\section{Numerical Results}
In this section, we will give simulations to further investigate the BER performance of our proposed SDMA-GrCIM scheme, and compare the simulations with the mathematical analysis in Section IV. The simulation parameters are set as follows:  if not specifically mentioned, the channel fading variances are all set to 1 for simplicity, and the common messages of all the UEs are the same (broadcast scenario). However, the same results can be obtained for unicast or multicast scenarios when the channel fading variances are other values as well.

\begin{figure}[ht]
  \centering
  \includegraphics[scale=0.6]{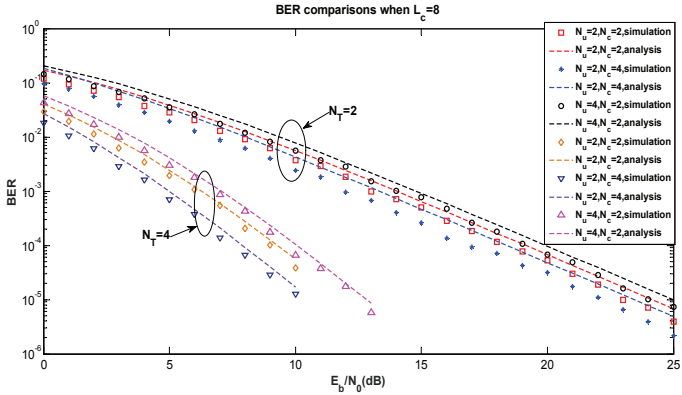}\\
  \caption{BER performance comparison with the length of Walsh codes set as $L_c=8$.}\label{Lc8}
\end{figure}

\begin{figure}[ht]
  \centering
  \includegraphics[scale=0.7]{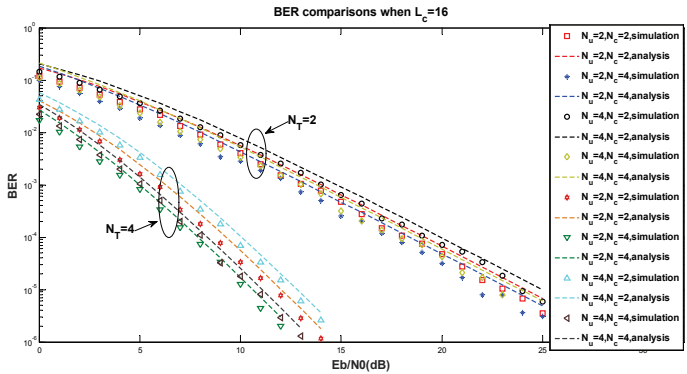}\\
  \caption{BER performance comparison with the length of Walsh codes set as $L_c=16$.}\label{Lc16}
\end{figure}

Fig.~\ref{Lc8} and Fig.~\ref{Lc16} depict the BER performance comparison with the lengths of Walsh codes are $L_c=8$ and $L_c=16$, respectively. Different numbers of transmit antennas and UEs are set for simulation to make comparisons. From the figures we can know that the numerical results of the BER coincide well with our mathematical analysis. Additionally, when all the other parameters are the same, the performance of more transmit antennas is better in BER. This is because the precoding operation provides beamforming gain actually, and more gains will be provided with larger number of transmit antennas. Besides, the numerical curves are more closer to the mathematical analysis ones for $N_T=4$ ($<0.5$dB) than that for $N_T=2$ ($<2$dB). This phenomenon can be explained through Eq. (\ref{PEP2}), where the omitted term $\Delta_n$ only is negligible in high SINR region.

The other conclusion is that BER performance with $N_c=4$ outperforms that with $N_c=2$ when all the other parameters are identical. The reason is that higher transmission rate leads to higher received SINR as illustrated in Section \ref{received-SINR}. However, if the variances of the UEs are the same, larger numbers of UE leads to degradation in BER performance, because the power coefficient $P$ will degrade as UE increases. This phenomenon does not always hold when the variances are not identical.

\section{Conclusion}
A rate splitting SDMA-GrCIM scheme is presented for 6G downlink transmission. The scheme exploits CIM technique to transmit the private and common messages, achieving the goal of rate splitting. Walsh codes are grouped into several orthogonal subsets to eliminate the inter-user interference. Mathematical upper bound BER is obtained for our scheme, and matches well with the numerical results. Non-grouped CIM scheme will be considered in our future work to improve the transmission rate.

\bibliographystyle{IEEEtran}
\bibliography{ref}
\end{document}